\documentclass[tighten,twocolappendix,apj]{openjournal}
\usepackage{tabularx}
\usepackage{graphicx}
\usepackage{cancel}
\usepackage{multirow}
\usepackage{amssymb}
\usepackage{bm}

\usepackage{amsmath}
\usepackage{comment}
\usepackage[pdftex, colorlinks=true, linkcolor=black, urlcolor=black,  citecolor=blue]{hyperref}
\usepackage{color}
\usepackage[pdftex, colorlinks=true, linkcolor=black, urlcolor=black,  citecolor=blue]{hyperref} 
\usepackage{subfigure} 
\usepackage{float}

\begin{document}
\title{Search for a distance-dependent Baryonic Tully-Fisher Relation at low redshifts }
\author{Aditi Krishak$^1$}
\email{aditikrishak@gmail.com}
\author{Shantanu Desai$^2$}
\email{shntn05@gmail.com}
\affiliation{$^{1}$ Department of Physics, IISER-Bhopal, Madhya Pradesh, 462066, India \\ $^{2}$Department of Physics, IIT Hyderabad, Kandi Telangana-502284, India}

\begin{abstract}
A recent work~\citep{AAP} has found a statistically significant transition in the  Baryonic Tully-Fisher relation  (BTFR) using low redshift data ($z<0.1$), with the transitions occurring at  about 9 and 17 Mpc. Motivated by this finding, we carry out a variant of this analysis by fitting the data to an  augmented  BTFR, where both  the exponent as well as normalization constant vary as a function of distance. We find that both the exponent and normalization constant show only a marginal variation with distance, and are consistent with a constant value,  to within $2\sigma$. We also checked  to see if there is a statistically significant difference between the BTFR results after bifurcating the dataset at distances of 9 and 17 Mpc. We find that almost all the sets of subsamples obey the BTFR with $\chi^2$/dof close to 1 and the best-fit parameters consistent across the subsamples. Only the subsample  with $D<17$ Mpc shows a marginal discrepancy (at  $1.75\sigma$)  with respect to the BTFR. Therefore, we do not find any evidence for statistically significant differences in the BTFR at distances of 9 and 17 Mpc.
\end{abstract}

\maketitle

\section{Introduction}
The Tully-Fisher relation is an empirical relation between the optical luminosity of spiral galaxies ($L$) and the asymptotic rotation velocity ($v$),  originally obtained using HI line widths, given by $L \propto v^4$~\citep{Tully}. This relation has been superseded by a related relation known as Baryonic Tully-Fisher relation (BTFR), where the luminosity is replaced by the total baryonic mass ($M_b$)~\citep{McGaugh00}
\begin{equation}
M_B=A v^4
\end{equation}
where $M_B$ is the sum of gas and star mass and $A$ is the normalization constant equal to $Ga_0$, where $a_0 \sim 10^{-10} m/s^2$. The BTFR has a much smaller scatter than the traditional Tully-Fisher relation~\citep{McGaugh2000,Marc,Zaritsky}, and  most of  its observed scatter can be attributed to observational uncertainties~\citep{Lelli15}.   The BTFR is known to be valid for a whole slew of galaxies besides spirals, such as low surface brightness and gas rich spiral galaxies, and it also works in cases where the classic T-F relation fails~\citep{Famaey12,McGaugh21}. A heuristic way of understanding the BTFR is by positing that the surface density ($M/R^2$) is constant~\citep{Huchra,Donato}. This surface density is known to be constant for a wide class of systems~\citep{Freeman,Donato,Salucci}, although there are exceptions in the form  of galaxy clusters and groups~\citep{Gopika1,Gopika2}.
The BTFR cannot be trivially predicted by the current $\Lambda$CDM model~\citep{Dolag} (although see~\citet{Aseem,Manoj}), but naturally follows as a consequence of Modified Newtonian dynamics~\citep{Famaey12,McGaugh14,Mcgaughgalaxies}. However, most relativistic theories of MOND have been ruled out after the simultaneous EM-GW observations from GW170817~\citep{Boran}.

Given the profound implications of the BTFR for new physics, there have been many theoretical~\citep{Sagi,Psaltis,Portinari} and observational~\citep{Ubler,Glowacki,AAP} (and references therein) efforts to search for  its evolution as a function of redshift. These searches have often led to contradictory results. However, all these works only compared the high redshift samples with the low redshift ones. There was no study exclusively devoted to the low redshift end of the BTFR. Motivated by these considerations, ~\citet{AAP} (A21, hereafter) analyzed the BTFR data at low redshifts ($z \lesssim 0.1$ ) corresponding to the distance range $\in [1,130]$ Mpc to search for sharp discontinuities in the slope and intercept. This could in turn be used to constrain the effective Newton's constant ($G_{eff}$). A21 considered a sample of 118 data points in the distance range 2$-$60 Mpc, collated from multiple sources~\citep{Ursa,Walter,Lelli15}. They analyzed the aforementioned data to  look for a transition in the evolution of BTFR, thereby constraining a possible evolution of $G_{eff}$, and to check if any putative variations can ameliorate the Hubble tension conundrum~\citep{Verde,Divalentino21, Bethapudi}.

A21 bifurcated the sample into two subsets based on a critical distance $D_c$. The BTFR was fit separately for each of the two  subsamples based on $\chi^2$  minimization, which includes an intrinsic scatter of 7.7\%. A21 found $3\sigma$ discrepancies  between the subsamples after the splitting the dataset at  $D_c$=9 Mpc and 17 Mpc. Such a transition could either be a systematic or a transition in the value of effective Newton's constant. If interpreted as a change in Newton's constant, A21 obtained $\frac{\Delta G_{eff}}{G_{eff}} \sim $ -10\%, where the minus sign corresponds to weaker gravity in the past. Such a magnitude of transition in $G$ is in accord with the model in ~\citet{Marra} designed to resolve the Hubble conundrum. However, the observed variation  is at odds with the bounds on variation of $G$~\citep{PDG,Bhagvati}. Note however that the best-fit BTFR parameters are still consistent between the two subsamples (See Table 2 of A21). They have also not considered the difference in number of data points between the two subsamples.

Nevertheless, in order to independently investigate the intriguing result in A21, we try to reproduce their claims using two independent ways. We first carry out a variant of this procedure where we search for a distance-dependent smooth variation in the BTFR.
We then try to characterize the differences in the subsamples after bifurcating the dataset at 9 and 17~Mpc.

This manuscript is organized as follows. We describe our analysis of the BTFR to search for distance-dependent variation in section~\ref{sec2}. In section~\ref{sec3}, we describe our analysis of the BTFR by splitting the dataset, with the intrinsic scatter as a free parameter as well as for a fixed value of the intrinsic scatter (section~\ref{sec3.1}). Finally, in section~\ref{sec4}, we summarize the results of our analyses and the conclusions that we draw from the results.

\section{Analysis of distance-dependent BTFR}
\label{sec2}
We start with the classical BTFR:
\begin{equation}
    M_B=Av^s
\end{equation}
This equation can be expressed in the logarithmic form as: 
\begin{equation}
\ln{M_B} = s\ln{v}+\ln{A}
\label{eq:btfr}
\end{equation}
or, $y=sx+b$, where $y \equiv \ln{M_B}$ and $x \equiv \ln{v}$. $s$ gives the slope and $b\equiv\ln{A}$ is the intercept of the logarithmic BTFR.

To search for distance-dependent transitions in the BTFR, we introduce a linear distance-dependence in the slope and intercept terms:
\begin{equation}
    y=(sD+c)x+bD+e
    \label{eq1}
\end{equation}
Thus, $s$ and $b$ encode the  distance-dependent slope and intercept, respectively, in our variant of this analysis.
The log-likelihood for the above equation can be expressed as~\citep{Gopika1,Pradyumna}
\begin{equation}
    -2\ln{\mathcal{L}}=\sum_i\ln(2\pi\sigma_i^2)+\sum_{i}\frac{\big[y_i-\big((sD_i+c)x_i+bD_i+e\big)\big]^2}{\sigma_i^2}
\end{equation} 
where the index $i$ runs over all data points in consideration. $\sigma_i^2$ incorporates uncertainties in observation of $x$, $y$, and $D$ along with the intrinsic scatter $\sigma_s$. Therefore,
\begin{equation}
    \sigma_i^2=\sigma_{y_i}^2+(sD_i+c)^2\sigma_{x_i}^2+(sx_i+b)^2\sigma_{D_i}^2 +\sigma_{s}^2
\end{equation}

Note that unlike A21, which fixed the scatter to a constant value of 7.7\%, we kept the intrinsic scatter ($\sigma_s$) as a free parameter. 
With the data points provided in A21, we perform parameter optimization using MCMC, using  the \texttt{emcee} sampler~\citep{emcee} for the model parameters $s$, $c$, $b$, and $e$. We used uniform priors across $\left[-100,100\right]$ for $s$, $c$, $b$, and $e$. A log-uniform prior between $10^{-5}$ and 1 was used for $\sigma_s$. The marginalized 68\% and 95\% credible interval contours for each of the free parameters were obtained using the {\tt getdist}~\citep{getdist} package. 

\begin{figure}[h]
    \centering
       \includegraphics[width=\columnwidth]{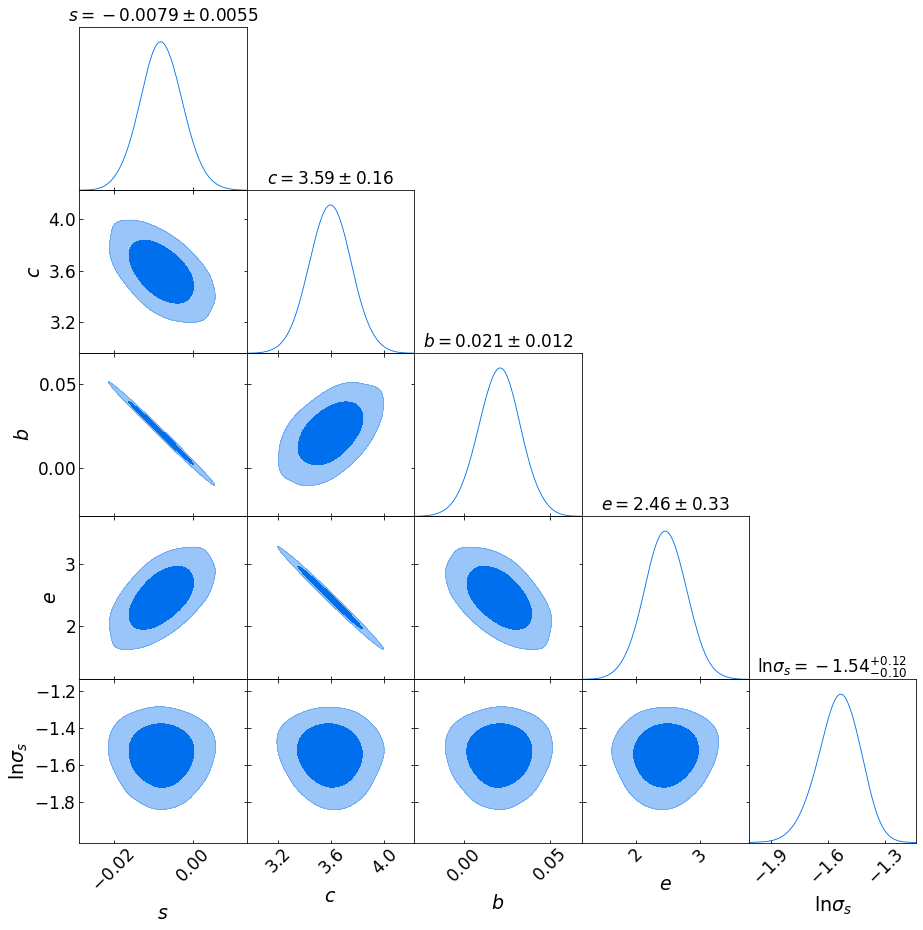} 
 \caption{The marginalized 68\% and 95\% credible intervals for all the free parameters in Eq.~\ref{eq1}. The contours have been produced  using the {\tt getdist} package in Python. \label{fig1}}
\end{figure}

\begin{figure}[h]
    \centering
       \includegraphics[width=\columnwidth]{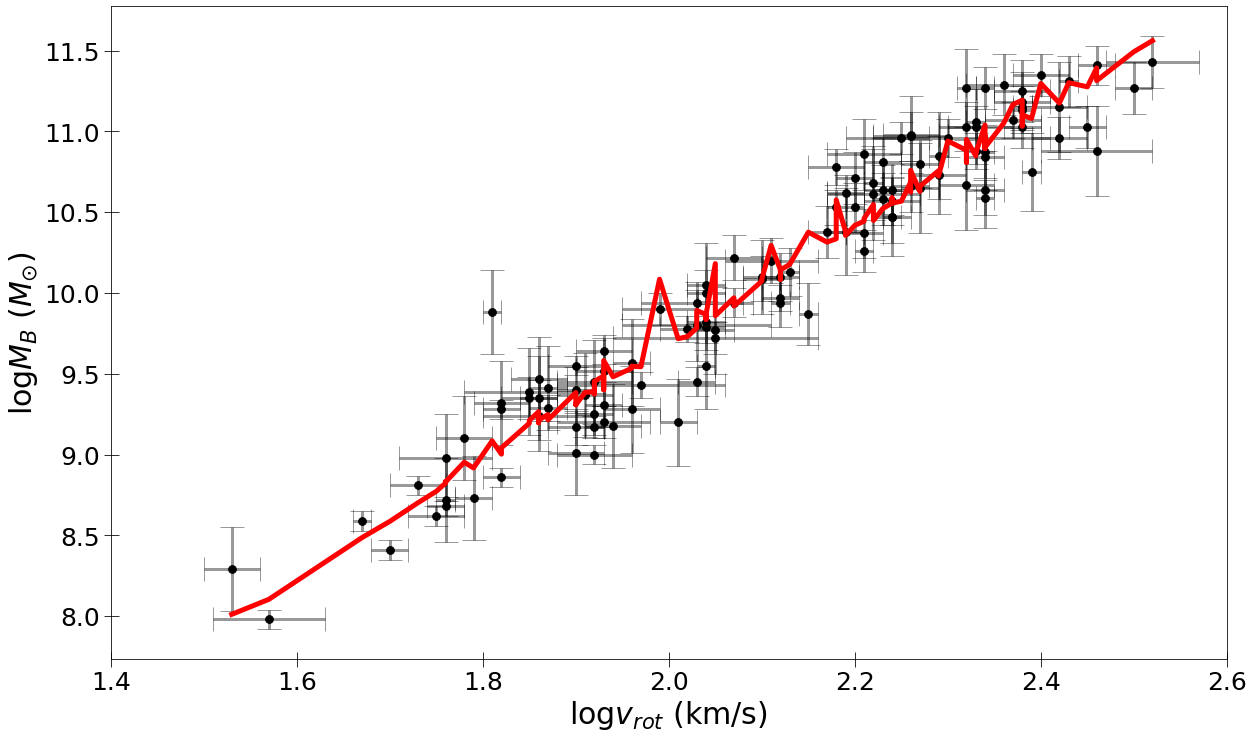} 
 \caption{The best-fit line for the BTFR (using the fits in Fig.~\ref{fig1}) along with the low-redshift data collated in A21~\label{fig2}. Note that the best-fit is also a function of distance which is not show in this figure. However, the distance varies for each data point and that is why the curve shows some bumps and wiggles. }
\end{figure}

The aforementioned marginalized credible intervals can be found in Fig.~\ref{fig1}.
We 
find that the best-fit values for $s$ and $b$ are given by $s=-0.0078 \pm 0.0054$ and $b=0.02 \pm 0.012$.  This shows that $s$ and $b$ are only marginally discrepant with respect to a constant value,  at $1.4\sigma$ and $1.6\sigma$, respectively.
Therefore, prima-facie there is no evidence for a distance-dependent slope or normalization in the BTFR, when considering low-redshift data.
We also find a strong degeneracy (anti-correlation) between $b$ and $s$, as well as between $e$ and $c$. The intrinsic scatter is given by  $\ln \sigma=-1.54 \pm 0.2\%$, corresponding to about  20\%. The best-fit values is shown on top of the data in Fig.~\ref{fig2}.

\begin{figure}
    \centering
\end{figure}

\section{Analysis of BTFR after splitting the dataset}
\label{sec3}
One possible reason for the discrepancy with respect to A21 is that our previous analysis may not be sensitive to a step function in the values of BTFR parameters at fixed distances. Therefore, instead of looking for a smooth distance-dependent BTFR parameters, we divide the dataset into two distinct subsamples  at distances  of  9~Mpc and 17 Mpc, since A21 had found about $3.5\sigma$ difference between the BTFR parameters at these bifurcation points.

For each of these subsamples, we fit the classical BTFR (Eq.~\ref{eq:btfr}) using the same Gaussian likelihood as was used for our distance-dependent BTFR fits. Again, we keep the intrinsic scatter as a free parameter. Our results for the different datasets can be found in Table~\ref{table1}. 
We can see that the best-fit values for the slope and intercept, (which correspond to the normalization and exponent in the BTFR, respectively) are consistent between the two subsamples at 1$\sigma$. The intrinsic scatter is also consistent across the subsamples.  Therefore, again we do not find any tension between the BTFR parameters  after bifurcating the dataset based on distances  corresponding to 9 and 17 Mpc.
The last two columns in Table~\ref{table1} show  the best-fit $\chi^2$ values and $p$-value,  where $\chi^2$ is calculated as follows:
\begin{equation}
\chi^2 = \sum_{i}\frac{\big[y_i-\big(sx_i+b\big)\big]^2}{\sigma_i^2} , 
\end{equation}
where $s$ and $b$ are the best-fit values for the slope and intercept ($b\equiv \ln A $), respectively. Although the $\chi^2$ between the two subsamples differ, the dof (degrees of freedom) are different because of the difference in the number of data points. Therefore, instead of looking at difference in the $\chi^2$ values between the subsamples as in A21, we consider the $\chi^2$ $p$-value for each of the subsamples~\citep{Desai16}. These $p$-values can be found in the last column of Table~\ref{table1}. All four subsamples 
have  $p$-value close to 1. the $\chi^2$/dof is also less than 1 for all subsamples. Therefore, there is no statistically significant differences between the subsamples.

 \begingroup
    \setlength{\tabcolsep}{10pt} 
    \renewcommand{\arraystretch}{1.75}
\begin{table}[]
    \centering
    \begin{tabular}{c|c|c|c|c|c}
    \hline
       \textbf{D (Mpc) }& \textbf{Slope} & \textbf{Intercept }& $\mathbf{\ln{\sigma_s}}$&$\mathbf{\chi^2/dof}$ & \textbf{p-value} \\
       \hline
       $<9$  & $3.38\pm 0.31$ &$2.84\pm 0.58$  & $-1.50_{-0.17}^{+0.21}$&14.61/29 & 0.988\\
       $>9$ & $3.38\pm 0.17$ &$3.05\pm 0.38$  &$-1.59_{-0.13}^{+0.16}$&43.71/83 & 1.000\\
       \hline
       $<17$ &$3.47\pm 0.18$  &$2.70\pm 0.35$ &$-1.48_{-0.13}^{+0.16}$&27.41/51 & 0.997\\
       $>17$ &$3.47\pm 0.21$  &$2.86\pm 0.46$ &$-1.66_{-0.15}^{+0.22}$&31.67/61 & 0.999\\
       
       \hline

    \end{tabular}
    \caption{Results from the best-fit BTFR parameters after bifurcating the sample at 9 Mpc (first two rows) and 17 Mpc (last two rows) with the intrinsic scatter as a free parameter. The best-fit parameters for both the slope and intercept are seen to be consistent between both the subsamples at 1$\sigma$. The $\chi^2/$dof for all four subsamples is less than one and $p-$ value is also close to one, indicating that there is no statistically significant differences across the samples.}
    \label{table1}
\end{table}
    \endgroup

\subsection{Analysis using a fixed intrinsic scatter} 
\label{sec3.1}
Finally,  we redo the analysis of BTFR (Eq.~\ref{eq:btfr}) using a fixed intrinsic scatter, instead of having it as a free parameter. We used the same value as A21 ($\sigma_{s}=0.0077$), which was obtained by positing that $\chi^2/dof$ was equal to one. The results from this  analysis can be found in Table~\ref{table2}. Once again, the best-fit parameters are consistent between the complementary subsamples within $1\sigma$. If we compare the datasets obtained after splitting the samples at 9 Mpc, we find that  the $\chi^2$/dof is close to 1 and $p$-value is also $> 0.1$ for all the subsamples.

On splitting the dataset at 17 Mpc, we find that the $p$-value sample with $D>17$ Mpc is close to one. However, the sample with $D<17$ Mpc has a $p$-value of 0.04,  equivalent to  a 1.75$\sigma$ discrepancy with respect to BTFR using the prescription in ~\citet{Cowan}. It is likely that there are additional systematics in this data with $D < 17$ Mpc that have not been accounted for, which is a possible reason for the small $p$-value.

\begingroup
    \setlength{\tabcolsep}{10pt} 
    \renewcommand{\arraystretch}{1.75}
\begin{table}[]
    \centering
    \begin{tabular}{c|c|c|c|c}
    \hline
       \textbf{D (Mpc) }& \textbf{Slope} & \textbf{Intercept }&$\mathbf{\chi^2/dof}$ & \textbf{p-value} \\
       \hline
       $<9$  & $3.50\pm 0.19$ &$2.61\pm 0.36$  & 40.12/30 & 0.102\\
       $>9$ & $3.45\pm 0.13$ &$2.88\pm 0.28$  & 84.70/84 & 0.458 \\
       \hline
       $<17$ &$3.57\pm 0.10$  &$2.51\pm 0.21$ & 70.68/52 & 0.043\\
       $>17$ &$3.57\pm 0.15$  &$2.64\pm 0.34$  & 53.57/62 & 0.768\\
       
       \hline

    \end{tabular}
    \caption{Results from the best-fit BTFR parameters after bifurcating the sample at 9 Mpc (first two rows) and 17 Mpc keeping the intrinsic scatter fixed at 0.077 similar to A21.}
    \label{table2}
\end{table}
    \endgroup
    
\section{Conclusions}
\label{sec4}
In a recent work, A21 used a dataset of 118 BTFR measurements at low  redshifts ($z<0.1$) to look for distance-dependent transitions in the parameters. They found a $3.7-4.5\sigma$ tension between the subsamples after splitting the data based on distance at 9 and 17 Mpc. This was ascertained based on the $\Delta \chi^2$ values of about 23.7 and 17, respectively. If this tension is interpreted as a change in the gravitational strength, it would imply a decrease in the effective Newton's constant of about 10\%, which has the right sign and order of magnitude needed to resolve the Hubble tension conundrum in Cosmology.

In order to independently test this intriguing result, we did two analyses. We augmented the standard BTFR by adding a distance-dependent term in both the normalization and exponent (cf. Eq.~\ref{eq1}). We fit  the data in logarithmic space to this distance-dependent BTFR. The best-fit values for the distance-dependent terms in both slope and intercept can be found in Fig.~\ref{fig1}. The best-fit values for the  distance-dependent slope ($s$) and intercept ($b$) are given by $s=-0.0078 \pm 0.0054$ and $b=0.02 \pm 0.012$. Therefore, these parameters show only a mild variation with distance and are consistent with no variation within 2$\sigma$.

Finally, we bifurcated the dataset into two subsamples based on the distance at 9 and 17 Mpc, followed by analysis of the usual BTFR for these subsamples. The best-fit values are consistent across the subsamples. When we kept the intrinsic scatter as free parameter, the $p$-value of the fit is close to 1. When we used the same fixed value for the intrinsic scatter as A21, we the $\chi^2$/dof for most of  the subsamples is close to one.  Only the subsample with $D<17$ Mpc shows a marginal discrepancy with the BTFR with $p$-value equal to 0.043, corresponding to a 1.75$\sigma$ discrepancy.  However, this discrepancy is marginal and not large enough to claim a statistically significant tension between the two subsamples.

Therefore, we conclude that we cannot corroborate the results in A21 giving $> 3\sigma$ tension between the subsamples at the critical  distances of 9 and 17 Mpc.

\bibliography{main.bib}
\end{document}